\documentclass[epj,twocolumn,a4paper]{svjour}
\usepackage{graphicx}
\usepackage{dcolumn}
\usepackage{bm}
\usepackage{parskip}
\usepackage{color}
\usepackage{subfigure}
\usepackage{amssymb}
\usepackage{slashed}
\usepackage[export]{adjustbox}
\usepackage{hyperref}
\usepackage{cite}
\usepackage{widetext}

\usepackage{amsmath}
\usepackage{amssymb}
\usepackage{graphicx}

\usepackage{feynmp-auto}
\usepackage{algorithm2e}
\usepackage{breqn}
\usepackage{mathtools}
\makeatletter
\newcommand*{\rom}[1]{\expandafter\@slowromancap\romannumeral #1@}
\makeatother

\title{Exploring high scale seesaw models through a supersymmetric portal}
\author{Yi Liu\inst{1} \and Stefano Moretti\inst{1,2} \and Harri Waltari\inst{2}}

\institute{School of Physics $\&$ Astronomy, University of Southampton, Highfield, Southampton, UK \\
\email{y.liu@soton.ac.uk, s.moretti@soton.ac.uk} \and Department of Physics $\&$ Astronomy, Uppsala University, Box 516, SE-75120 Uppsala, Sweden\\
\email{stefano.moretti@physics.uu.se, harri.waltari@physics.uu.se}}

\date{Received: date / Revised version: date}

\abstract{
The seesaw scale is a priori unknown. If the seesaw scale is low, one may directly look for the new particles predicted by seesaw models. If the seesaw scale is high, such an approach is unfeasible. We show that in some supersymmetric seesaw models the large Yukawa couplings of high scale seesaw models leave their fingerprints to Higgs-slepton couplings and that this can result in decays of the type $\tilde{\nu}_{2}\rightarrow \tilde{\nu}_{1}h$ in Type-I and Type-III seesaw models and $\tilde{\ell}^{\pm}_{2}\rightarrow \tilde{\ell}^{\pm}_{1}h$ in the latter. Unfortunately the current exclusion bounds make it impossible to see a significant signal even at the High-Luminosity phase of the LHC. In this paper, we highlight that the High-Energy phase of the LHC (with $\sqrt{s}=27$~TeV) could afford one with some sensitivity to those in the single lepton channel.
}


\begin{document}

	\maketitle

\newpage

\section{Introduction}

Neutrino masses have been known to be non-zero for 25 years \cite{Super-Kamiokande:1998kpq}. As they are so much smaller than all other Standard Model (SM) fermion masses, one usually assumes that they are generated by some kind of a seesaw mechanism \cite{Minkowski:1977sc,Konetschny:1977bn,Gell-Mann:1979vob,Mohapatra:1980yp,Foot:1988aq}. The masses are still generated through the Higgs mechanism, but suppressed by a heavy seesaw particle, which can be a singlet neutrino (Type-I), a triplet of Higgs bosons (Type-II) or a triplet of exotic leptons (Type-III) (see Refs. \cite{Khalil:2022toi,Moretti:2019ulc} for reviews).

The seesaw scale is a priori unknown. If the seesaw scale is around the Electro-Weak (EW) scale, one may be able to produce the seesaw particles directly at the Large Hadron Collider (LHC) \cite{CMS:2017ybg,CMS:2018jxx,ATLAS:2019kpx,ATLAS:2020wop}. One of the original ideas \cite{Gell-Mann:1979vob} was that the smallness of the  neutrino masses could be related to the breaking of a Grand Unification Theory (GUT), {\it i.e.}, the relevant Yukawa couplings would be of  order unity and the seesaw scale somewhere around $\alpha M_{\mathrm{GUT}} \sim 10^{14}$~GeV. Such energy scales are obviously out of the reach of present and future colliders.

Supersymmetry, the symmetry between fermions and bosons, is often a necessary ingredient in formulating models with large separations of scales. Due to the cancellation between the bosonic and fermionic loops, the separation of scales is radiatively stable \cite{Dimopoulos:1981zb}, once it has been generated by some dynamics. Thus in the supersymmetric framework, scalar masses would not get quadratic corrections proportional to the seesaw scale and an EW scale Higgs boson would not be unnatural even if the seesaw scale was close to the GUT scale.

In the context of high scale seesaw models, supersymmetry has one remarkable property. The scalar potential, and especially its $F$-terms being of the form
\begin{equation}
V=\sum_{i} \left| \frac{\partial W}{\partial \varphi_{i}} \right|^{2},
\end{equation}
leads to four-scalar interactions without the seesaw particle but with the seesaw couplings involved. If the couplings are of the order unity, they are among the largest ones in the model and could lead to observable consequences.

For definiteness, let us consider the Type-I seesaw model, where the extra superpotential terms in addition to those of the Minimal Supersymmetric Standard Model (MSSM) are
\begin{equation}
    W=W_{\mathrm{MSSM}}+ y^{\nu} L\cdot H_{u} N^{c}+M_{N}N^{c}N^{c},
\label{eq:seesaw1}
\end{equation}
where we assume $y^{\nu}\sim 1$ and $M_{N}\sim 10^{14}$~GeV. When differentiating with respect to $N^{c}$, one gets the term $$\sum_{k} y^{\nu *}_{ik}y^{\nu}_{jk}\tilde{L}^{\dagger}_{i}\cdot H_{u}^{\dagger}\tilde{L}_{j}\cdot H_{u},$$ involving only Higgs bosons and left-handed sleptons, which we assume to be at the TeV scale. If there are significant mass splittings between the sfermion generations, which could well be generated through Renormalisation Group Evolution (RGE) due to the large couplings, one might get processes like $\tilde{\nu}_{i}\rightarrow \tilde{\nu}_{j}h$ with a large Branching Ratio (BR). If the sneutrinos decay visibly, the decays can be distinguished from mono-Higgs signatures that could arise from dark matter  \cite{Petrov:2013nia,Berlin:2014cfa}. Slepton decays with Higgs bosons in the final state could offer an indication of a high scale seesaw model and thus provide us a window to scales otherwise beyond our experimental reach.

Our aim is to investigate how could one observe such slepton decay patterns involving Higgs bosons in seesaw models of Type-I and Type-III, which have a similar structure in terms of the TeV scale Lagrangian.

Our paper is organised as follows. Higgs-slepton interactions are described in the next section, which is followed by a discussion of the production and decay modes relevant to our research. Our numerical analysis is introduced in the following section, after which we conclude.

\section{Higgs-slepton interactions in seesaw models}

We shall now look at how the Higgs-slepton interactions arise from our seesaw models in some detail. In particular, we look at Type-I and Type-III seesaw models. Both have Yukawa couplings that connect the lepton and Higgs doublets to the seesaw particles, which form a singlet and triplet under SU(2). The superpotential of Type-I seesaw is given in Eq. (\ref{eq:seesaw1}) and for Type-III seesaw it is

\begin{equation}
    W = W_{\rm{MSSM}} + y^\nu L \Sigma H_u + M_\Sigma \mathrm{Tr}(\Sigma^{2}),
\end{equation}
where $L$ is the left-chiral lepton doublet and $H_u = (H^+ , H^0)^T$ is the up-type Higgs doublet. The $\Sigma$ is an antilepton ($L=-1$) chiral superfield which transforms as $(1,3,0)$ under the SM gauge group $SU(3)_c\times SU(2)_L \times U(1)_Y$. The mass term for $\Sigma$ violates lepton number by two units.

The superfield $\Sigma$ can be represented 
\begin{equation}
    \Sigma = \sigma^{i}\Sigma^{i}= \Big ( \begin{array}{cc}
       \Sigma^0/\sqrt{2}  &   \Sigma^+ \\
        \Sigma^- & -\Sigma^0/\sqrt{2}
    \end{array}\Big), \quad \Sigma^\pm = \frac{\Sigma^1 \mp i\Sigma^2}{\sqrt{2}}, \quad \Sigma^0 = \Sigma^3.
\end{equation}

The models look very similar in what comes to neutrino mass generation, both having a lepton and a Higgs doublet coupling to the companion neutrinos. The only difference is that the $L$ and $H_{u}$ superfields combine to a singlet in the case of Type-I and to a triplet in the case of Type-III seesaw. This difference between the two seesaw models leads to a difference in the scalar potential which contributes the processes that lead to slepton decays containing a Higgs boson.

When we expand the neutrino Yukawa terms in the superpotential, we get

\begin{widetext}
\begin{eqnarray}
    W  & = & y^{\nu}_{ij} \left( e^{-}_{i}H_{u}^{+}-\frac{1}{\sqrt{2}}\nu_{i} H_{u}^{0}\right)N^{c}_{j} +\ldots,\\
    W  & = & y^{\nu}_{ij} \left( \frac{1}{\sqrt{2}}e^{-}_{i}H_{u}^{+}\Sigma^{0}_{j} -\nu_{i}\Sigma^{-}_{j}H_{u}^{+}+\frac{1}{\sqrt{2}}e^{-}_{i}\Sigma^{+}_{j}H_{u}^{0}+\frac{1}{2}\nu_{i}\Sigma^{0}_{j}H_{u}^{0}\right)+\ldots,    
\end{eqnarray}
\end{widetext}
for Type-I and Type-III, respectively. Here we have included a factor of $1/\sqrt{2}$ into the definition of the neutral Higgs field.

Differentiating with respect to the heavy seesaw fields leads to the scalar potentials
\begin{eqnarray}
    V & = & \sum_{k}\frac{1}{2}y^{\nu}_{ik}y^{\nu *}_{jk}\tilde{\nu}_{i}\tilde{\nu}^{*}_{j}H_{u}^{0}H_{u}^{0 *}+\ldots, \label{eq:scalarpot1}\\
    V & = & \sum_{k}\frac{1}{4}  y^{\nu}_{ik}y^{\nu *}_{jk}\left(\tilde{\nu}_{i}\tilde{\nu}^{*}_{j}H_{u}^{0}H_{u}^{0 *}+2\tilde{e}^{-}_{i}\tilde{e}^{+}_{j}H_{u}^{0}H_{u}^{0 *}\right)+\ldots ,\label{eq:scalarpot3}
\end{eqnarray}
for Type-I and Type-III, respectively. Hence one in general gets Higgs interactions with sleptons that are non-diagonal in flavour space and, in the case of a high scale seesaw, have large couplings. After EW Symmetry Breaking (EWSB) we have $\langle H_{u}^{0}\rangle = v\sin\beta$ ($v=246$~GeV), which generates a three-point coupling between sleptons and the SM-like Higgs. 

One may also note that in Type-III seesaw there is a non-flavour-diagonal coupling between charged sleptons and Higgs bosons, while there is no such coupling in the case of Type-I seesaw. As we discuss below, this leads to a stronger signal arising from Type-III than Type-I seesaw. We further notice that, while the usual $D$-terms of the scalar potential also contain large couplings between sneutrinos, charged sleptons and Higgs bosons, such couplings are always flavour-diagonal and cannot result in decays of the type $\tilde{\nu}_{2}\rightarrow \tilde{\nu}_{1}h$, which is our smoking gun signature for high scale seesaw models.

Besides the decay modes containing Higgs bosons, there are other decay channels and the visibility of the signal depends on the branching ratios. If the Lightest Supersymmetric Particle (LSP) is a higgsino-like neutralino and the gauginos are heavier than the sleptons, the decays of the left-handed sleptons arise from the superpotential term $y^{\ell} LH_{d}E^{c}$, so one gets the decays $\tilde{\nu}\rightarrow \tilde{\chi}^{\pm}\ell^{\mp}$ and $\tilde{\ell}^{\pm}\rightarrow \tilde{\chi}^{0}\ell^{\pm}$. These lead to partial widths
\begin{eqnarray}
    \Gamma(\tilde{\nu}_{j}\rightarrow \ell^{\pm}_{j}\tilde{\chi}^{\mp}_{i}) & = & |y^{\ell}_{jj}|^{2}|U_{i2}|^{2}\frac{(m_{\tilde{\nu}}^{2}-m_{\tilde{\chi}}^{2})^{2}}{32\pi m_{\tilde{\nu}}^{3}},\label{eq:parwidth1} \\
    \Gamma(\tilde{\ell}^{\pm}_{j}\rightarrow \ell^{\pm}_{j}\tilde{\chi}^{0}_{i}) & = & |y^{\ell}_{jj}|^{2}|N_{i3}|^{2}\frac{(m_{\tilde{\ell}}^{2}-m_{\tilde{\chi}}^{2})^{2}}{16\pi m_{\tilde{\ell}}^{3}},\label{eq:parwidth2}
\end{eqnarray}
where $U_{i2}$ gives the higgsino component of the chargino (for our benchmarks $|U_{i2}|\simeq 1$), $N_{i3}$ gives the down-type higgsino component of the neutralino (for our benchmarks $|N_{13}|\simeq 1/\sqrt{2}$). If the soft slepton masses are not flavour diagonal, an appropriate linear combination of the leptonic Yukawas corresponding to the flavour composition of the sleptons must be used.

If the LSP is a gaugino there are additional decay channels $\tilde{\nu}\rightarrow \nu\tilde{\chi}^{0}$ and $\tilde{\ell}^{\pm}\rightarrow \tilde{\chi}^{\pm}\nu$ (if winos are light) and the decay widths are propotional to $g^{2}$ instead of $|y^{\ell}|^{2}$ and gaugino components instead of higgsino components. Since we have the hierarchy $y^{\ell}_{11}\ll y^{\ell}_{22} \ll y^{\ell}_{33} \ll g$, the strength of our signal will depend on the nature of the light neutralinos and charginos and in the case of higgsinos, the flavour of the heavier sleptons. As the electron and muon Yukawas are so tiny, in practice the mixing between the gaugino and higgsino components will be significant for the overall decay widths of the sneutrinos and charged sleptons unless the gauginos are extremely heavy.

We shall concentrate on the higgsino case, since as we shall see, already the tau Yukawa is so large that the signal containing Higgs bosons will have a too small branching ratio if stau is the heavy slepton that decays. Hence in all our benchmarks we make our gauginos heavier than the sleptons.

\section{The production and decay mechanisms}

To study the high-scale seesaw signatures with Higgs bosons, we build some Benchmark Points (BPs) with $m(\tilde{e}^{\pm})<m(\tilde{\mu}^{\pm})<m(\tilde{\tau}^{\pm})$ and mass splittings between generations larger than $m_{h}\approx 125$ GeV (the mass of the SM-like state $h$). As we shall see, this will be the limiting case, where we still can see a signal. If the second slepton (assuming the third one to be too heavy to be produced efficiently) would be a selectron, the signal would be similar (as the mixing with gauginos dominates the other decay modes already for smuons), while in the case of a stau, the signal would almost vanish due to the larger partial widths from equations (\ref{eq:parwidth1}) and (\ref{eq:parwidth2}). We consider the charged current process $pp\to \tilde{\ell}_{2}^{\pm}  \tilde{\nu}_2$, where the subscript indicates mass ordering. The charged current portal is more promising as the final state contains charged leptons even when the sneutrino decays invisibly.

\begin{figure*}[!hbt]
    \centering
    \begin{fmffile}{Seesaw3}
  \begin{fmfgraph*}(300,150)
    \fmfset{arrow_len}{3.5mm}
    \fmfstraight
    \fmfleft{i1,i2,i3,i4,i5}
    \fmfright{o1,o2,o3,o4,o5,o6,o7}
    \fmf{fermion,tension=3}{i2,v1,i4}
    \fmf{boson,label={$W^{\pm}$},tension=5}{v2,v1}
    \fmf{phantom,tension=1}{o2,v2,o6}
    \fmflabel{$q$}{i2}
    \fmflabel{$\overline{q}^{\prime}$}{i4}
    \fmffreeze
    \fmf{dashes,tension=0.5}{v2,v3}
    \fmf{dashes,tension=0.5}{v2,v4}
    \fmf{phantom,label={$\Tilde{\mu}^\pm$},tension=0,label.side=right}{v2,v4}
    \fmf{phantom,label={$\Tilde\nu_2$},tension=0,label.side=left}{v2,v3}
    \fmf{phantom,tension=1}{o1,v4,o4}
    \fmf{phantom,tension=1}{o1,v6,o2}
    \fmf{phantom,tension=1}{o5,v3,o7}
    \fmf{phantom,tension=1}{i2,v4,v3,i4}
    \fmffreeze
    \fmf{fermion,tension=2,label={$\Tilde{\chi}^\pm_1$},label.side=left}{v3,v5}
    \fmf{fermion}{v5,o7}
    \fmf{boson}{v5,o6}
    \fmf{fermion}{o5,v3}
    \fmflabel{$\Tilde{\chi}_0$}{o7}
    \fmflabel{$W^\pm$}{o6}
    \fmflabel{$\mu^\mp$}{o5}
    \fmf{dashes,tension=2,label={$\Tilde{e}^\pm$},label.side=right}{v4,v6}
    \fmf{dashes,tension=2}{v4,o4}
    \fmflabel{$h$}{o4}
    \fmf{fermion,tension=1.4,label={$\Tilde{\chi}_1^\pm$},label.side=right}{v6,v7}
    \fmf{fermion,tension=0.5}{v6,o3}
    \fmflabel{$\nu_e$}{o3}
    \fmf{fermion}{v7,o1}
    \fmf{boson}{v7,o2}
    \fmflabel{$W^\pm$}{o2}
    \fmflabel{$\Tilde{\chi}_0$}{o1}
    \fmfshift{30 right}{o1,o2}
    \fmfshift{40 left}{v6}
    \fmfshift{30 right}{o3}
  \end{fmfgraph*}
\end{fmffile}
    \caption{The dominant process for charged current slepton-sneutrino production and the subsequent decay involving a Higgs boson in Type-III seesaw. The $W^+W^-$ pair of gauge bosons will decay into leptons and/or jets.
    \label{fig:seesaw3}}

\end{figure*}
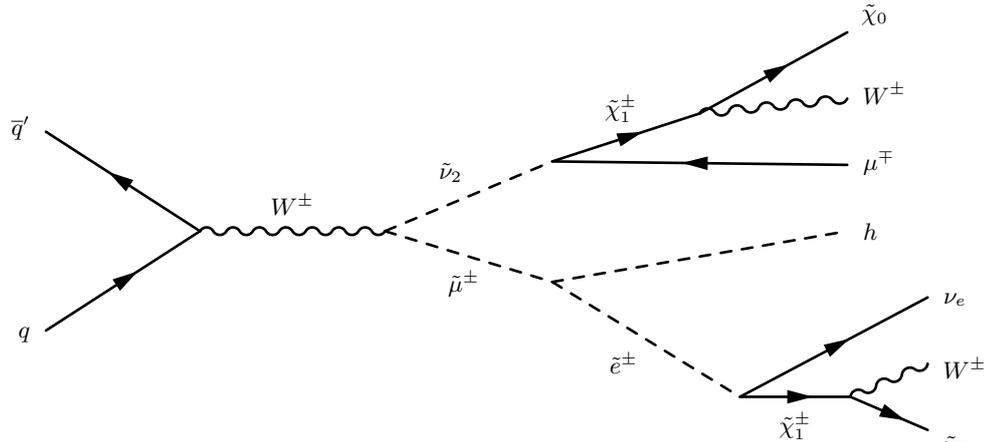

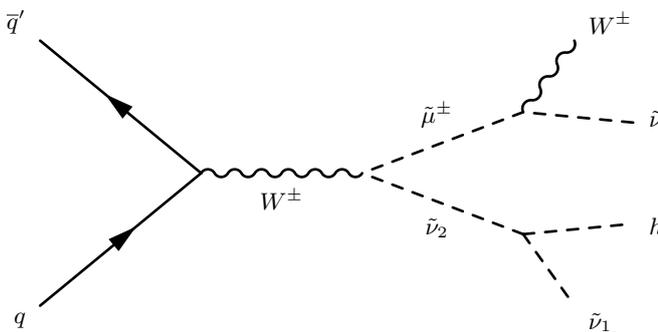
\begin{figure}[h]
\centering
\begin{fmffile}{s1}
 \begin{fmfgraph*}(250,100)
   \fmfleft{i1,i2,i3,i4}
   \fmfright{o1,o2,o3,o4}
   \fmf{fermion}{v1,i4}
   \fmf{fermion}{i1,v1}
   \fmflabel{$q$}{i1}
   \fmflabel{$\overline{q}^{\prime}$}{i4}
   \fmf{boson, tension=2, label={$W^{\pm}$}}{v1,v2}
   \fmf{dashes, label={$\tilde\mu^{\pm}$}, label.side=right}{v4,v2}
   \fmf{dashes, label={$\tilde\nu_2$}, label.side=left}{v3,v2}
   \fmf{dashes}{v3,o2}
   \fmf{dashes}{v3,o1}
   \fmf{boson}{o4,v4}
   \fmf{dashes}{o3,v4}
   \fmflabel{$W^\pm$}{o4}
   \fmflabel{$\tilde{\nu}_1$}{o3}
   \fmflabel{$\tilde{\nu}_1$}{o1}
   \fmflabel{$h$}{o2}
 \end{fmfgraph*}
\end{fmffile}
    \caption{The dominant process for charged current slepton-sneutrino production and the subsequent decay involving a Higgs boson in Type-I seesaw. The $W^+W^-$ pair of gauge bosons will decay into leptons and/or jets.}
    \label{seesaw1}
\end{figure}

As discussed above, in Type-III seesaw both sneutrinos and charged sleptons can decay to final states with Higgs bosons. The dominant process is $\tilde{\ell}_{2} \to \tilde{\ell}_{1} h$ while $\tilde\nu_2 \to \ell^{\pm} \tilde{\chi}_{1}^{\mp}, \nu \tilde{\chi}^{0}$. The Feynman diagram for such a process is shown in Fig. \ref{fig:seesaw3}. There is also a process, where the Higgs originates from a sneutrino decay, but that has a smaller BR as can be seen from equation (\ref{eq:scalarpot3}). In Type-I seesaw, only the sneutrino can decay into a Higgs boson via $\tilde\nu_2 \to h  \tilde\nu_1$. The corresponding Feynman diagram is shown in Fig. \ref{seesaw1}.

These processes can lead to a variety of final state topologies. Currently the limit for charged slepton masses is $m(\tilde{e}^{\pm}),m(\tilde{\mu}^{\pm})> 700$~GeV for neutralino masses below $350$~GeV \cite{ATLAS:2019lff}, which we take as our lower limit of charged slepton masses\footnote{With more compressed spectra $m(\tilde{\ell})-m(\tilde{\chi}^{0})\lesssim 100$~GeV, one obviously can have significantly lighter sleptons. Such cases need a different analysis strategy than the one  adopted here as we rely on large $\slashed{E}_{T}$ to suppress SM backgrounds.}. This means that the overall production rate of slepton-sneutrino pairs will be low, especially as we have to produce second generation sleptons with a large mass splitting compared to the first generation ones.

In fact, the production rate at the LHC even with nominal collision energy ($\sqrt{s}=14$~TeV) is so low ($\sim 30$~ab for $1$~TeV sleptons), that there will not be sufficient statistics even at the High-Luminosity LHC (HL-LHC) \cite{Gianotti:2002xx}. Hence we turn to the proposed High-Energy LHC (HE-LHC) \cite{FCC:2018bvk} with a nominal collision energy of $\sqrt{s}=27$~TeV. This increases the production cross section by an order of magnitude compared to the standard LHC.

\begin{table}[hbt]
    \centering
    \begin{tabular}{|c|c|c|}
    \hline
         & Type-III & Type-I \\ \hline
        $N(\ell)=0$ & 2484 & 1842 \\ \hline
        $N(\ell)=1$ & 2816 & 2940 \\ \hline
        $N(\ell)=2$ & 794  & 1268 \\ \hline
        $N(\ell)\geq 3$ & 58   & 169 \\ \hline
    \end{tabular}
    \caption{Lepton number multiplicities for typical BPs in both seesaw models. The luminosity is 10 ab$^{-1}$ and the energy is $\sqrt{s}=27$~TeV.}
    \label{tab:Nlep}
\end{table}

In Tab. \ref{tab:Nlep} we show the lepton multiplicities for some typical benchmark points (BP1 and BP3, defined in Table \ref{Benchmarks}). We see that the single lepton final state has the highest multiplicity for both seesaw models. As we will lose a part of the signal due to different BRs involved in the model, it is reasonable to look at the state with the highest multiplicity first. We also pick the Higgs decay mode to $b$-quarks as that has the highest BR and allows to reconstruct the Higgs boson, although not with a too high precision in mass. Unfortunately the channels with good mass resolution ({\it i.e.}, $\gamma\gamma$ and $ZZ^{*}\to 4$~leptons) are too rare to be useful with such a small event rate.

Our signal events will then consist of events with a single lepton, two $b$-tagged jets and missing momentum carried by the LSP. The largest SM backgrounds to this final state arise from the following processes:
\begin{itemize}
    \item $t\bar{t}$ production where one the top (anti)quarks decays semileptonically and the other one hadronically; 
    \item $W^{\pm}h$ production in the case where the $W^{\pm}$ boson decays into a lepton and a neutrino.
\end{itemize}
These have been considered to be the dominant backgrounds in similar types of experimental analyses (\textit{e.g.}, \cite{ATLAS:2022enb}).

\section{Simulation and results}
In this section we will describe our numerical toolbox and the Monte Carlo (MC) simulations that we have pursued with it.
\subsection{Analysis strategy}
The model files are produced by the Mathematica package \textsc{Sarah} v4.14 \cite{Staub:2015kfa}. This code also generates a source code for \textsc{Spheno} v4.0.4 \cite{Porod:2003um,Porod:2011nf} to obtain the mass spectrum and couplings as well as for \textsc{Madgraph5} v2.8.2 \cite{Alwall:2011uj} to simulate collider events. We use \textsc{Pythia} v8.2 \cite{Sjostrand:2014zea} for parton showering and hadronisation while we simulate the detector response by using \textsc{Delphes3} \cite{deFavereau:2013fsa}. We simulate the analysis and present our numerical results with \textsc{Madanalysis5} v1.8 \cite{Conte:2012fm}.

We prepare two BPs for Type-III seesaw and two for Type-I seesaw, which can be detected in the HE-LHC with 27 TeV collision energy and the integrated luminosity $10$~ab$^{-1}$. We simulate proton-proton collisions to produce the second generation sneutrino ($\tilde\nu_2$) and slepton ($\Tilde{\ell}_2$), which in our cases are smuon-like, and select decays to the SM-like Higgs boson plus corresponding first generation particles. The mass of $\tilde\nu_2$ and $\Tilde{\ell}_2$ should be heavy enough to allow for the decay kinematics. At the same time, the mass of lightest slepton is required to be larger than 700 GeV \cite{ParticleDataGroup:2022pth}. The particle mass spectra and relevant BRs are shown in Tab. \ref{Benchmarks}. 

\begin{table*}[h]
    \centering
    \begin{tabular}{|c|c|c|c|c|}
    \hline
    & BP1 & BP2 & BP3 & BP4 \\
    \hline
    $m(\Tilde{\mu})$ (GeV) & 895.3 & 1001.9 & 885.5 & 993.2 \\
    $m(\Tilde{e})$ (GeV) & 701.9 & 834.0 & 692.9 & 993.2\\
    $m(\tilde\nu_2)$ (GeV) & 886.8 & 994.3 & 891.7 & 998.6\\
    $m(\tilde\nu_1)$ (GeV) & 692.8 & 826.2 & 697.5 & 830.2\\
    $\mathrm{BR}(\Tilde{\ell}_{2}\rightarrow \Tilde{\ell}_{1}+h)$ & $74.6\%$ & $63.1\%$  & $0\%$ & $0\%$ \\
    $\mathrm{BR}(\tilde\nu_2\rightarrow \tilde\nu_1+h)$ & $12.2\%$ & $6.7\%$ & $30.9\%$ & $20.4\%$ \\
    LSP & 410.4($\tilde{\chi}_0$) & 410.4($\tilde{\chi}_0$) & 410.4($\tilde{\chi}_0$) & 410.4($\tilde{\chi}_0$) \\
    NLSP & 413.3 ($\tilde{\chi}_{1}^{\pm}$) & 413.3($\tilde{\chi}_{1}^{\pm}$) & 413.3 ($\tilde{\chi}_{1}^{\pm}$) & 413.3 ($\tilde{\chi}_{1}^{\pm}$) \\
    \hline
    \end{tabular}
    \caption{Mass spectra and BRs of our BPs. BP1 and BP2 are for Type-III seesaw while BP3 and BP4 are for Type-I seesaw. For all of these benchmarks we have bino and wino-like neutralinos with masses of $1100$~GeV and $2300$~GeV, which leads to a gaugino component of about $0.3\%$ for the LSP. The relevant neutrino Yukawas are in the range $0.3<|y^{\nu}|<0.5$ with $\sum_{k}y^{\nu}_{1k}y^{\nu}_{2k}\simeq 0.2$.}
    \label{Benchmarks}
\end{table*}

All of the BPs have the same Lightest Supersymmetric Particle (LSP) and Next-to-LSP (NLSP), which are higgsino-like neutralinos and charginos. BP1 has a mass spectrum similar to BP3 and the same situation arises between BP2 and BP4. However, there is a significant difference in the Higgs production cross section times BRs between Type-III seesaw and Type-I seesaw. For the sneutrino decay process, Type-I seesaw has  BRs larger than the Type-III ones, which can be traced back to the factors in equations (\ref{eq:scalarpot1}) and (\ref{eq:scalarpot3}). However, the charged slepton decay channel does not exist in Type-I seesaw whereas it dominates the Higgs signal in Type-III seesaw, consistent with equations (\ref{eq:scalarpot1}) and (\ref{eq:scalarpot3}).  As the slepton masses increase, the BR shows a decreasing trend. 

The BR for $\tilde{\mu}^{\pm}\rightarrow \tilde{e}^{\pm}h$ is high in Type-III seesaw, since the competing decay mode of eq. (\ref{eq:parwidth2}) is proportional to the small muon Yukawa coupling squared or the small gaugino-higgsino mixing factor squared. Had the second slepton been a selectron, the BR would have been similar as the gaugino-higgsino mixing would dominate the decays to neutralinos/charginos, while for staus the corresponding branching ratio is only a few percent as the tau Yukawa is large enough to dominate the branching ratio.

\begin{table*}[h]
    \centering
    \begin{tabular}{|c|c|}
    \hline
        Number of leptons & $N(\ell)=1$ \\
    \hline
        Number of $b$-jets & $N(b) \geq 2$ \\
    \hline
    Transverse momentum of electron & $p_T(e) >$ 5 GeV \\
    \hline
    Transverse momentum of muon & $p_T(\mu) >$ 3 GeV \\
    \hline
    Transverse momentum of jet & $p_T(j) >$ 17 GeV \\
    \hline
    Absolute pseudorapidity of electron & $|\eta(e)| < $ 2.5 \\
    \hline
    Absolute pseudorapidity of muon & $|\eta(\mu)| < $ 2.4 \\
    \hline
    Absolute pseudorapidity of jet & $|\eta(j)| < $ 2.4 \\
    \hline
    Relative distance  in  ($\eta,\phi$) between any two visible objects & $\Delta R \geq 0.5$ \\
    \hline
    \end{tabular}
    \caption{The multiplicity requirements for our final state topology ($\ell = e,\, \mu$), and the definitions of the various objects. Note that we allow for any number of non $b$-jets. 
  }
    \label{pre-selection}
\end{table*}

\begin{figure}
    \centering
    \includegraphics[scale=0.27]{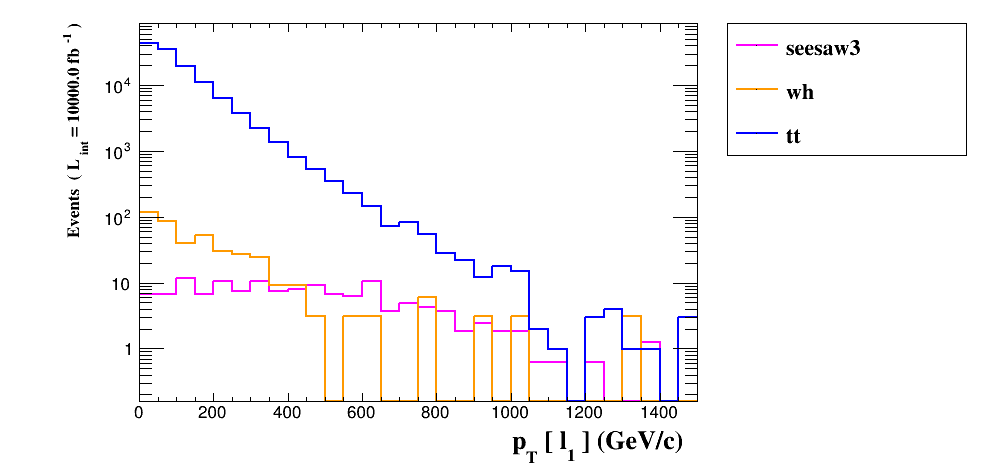}
    \caption{The distribution in transverse momentum of the leading lepton after the pre-selection (and generation-level) cuts.}
    \label{PTL1}
\end{figure}

\begin{figure}
    \centering
    \includegraphics[scale=0.27]{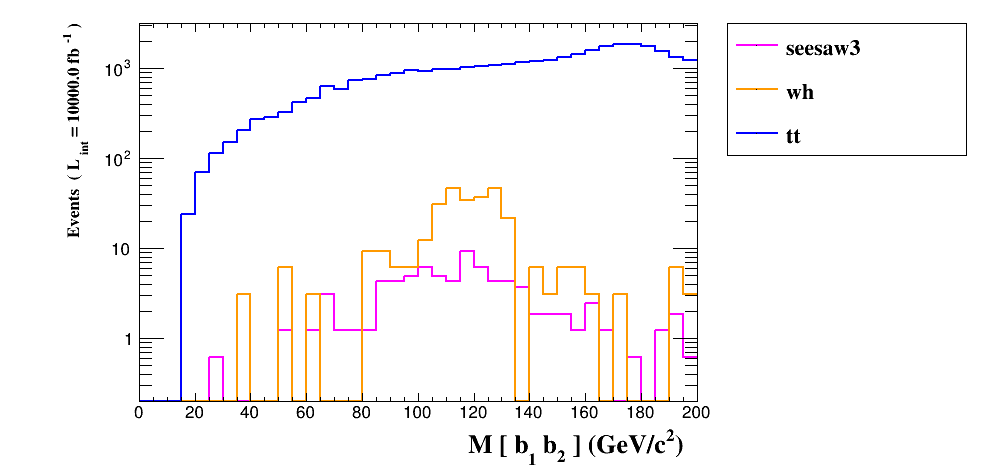}
    \caption{The distribution in invariant mass of the leading and next-to-leading $b$-jets after the pre-selection (and generation-level) cuts.}
    \label{IMbb}
\end{figure}

\begin{figure}
    \centering
    \includegraphics[scale=0.27]{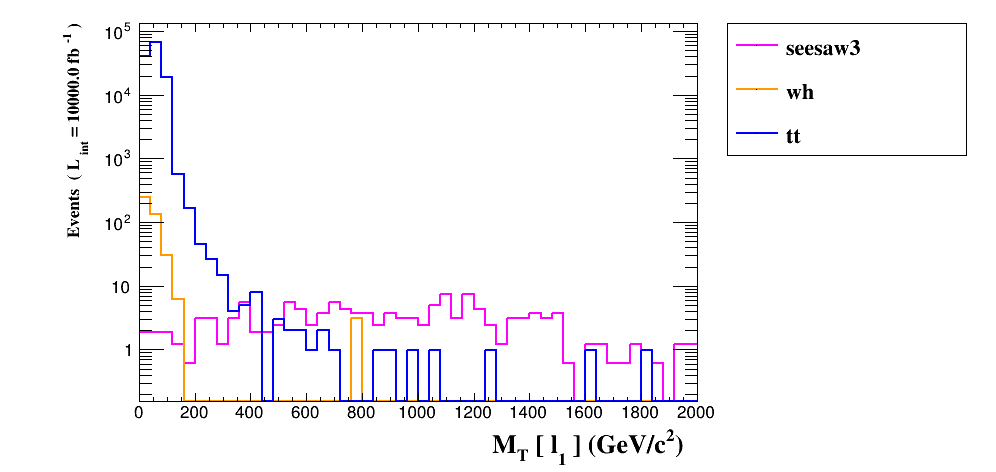}
    \caption{The distribution in transverse mass of leading lepton plus MET after the pre-selection (and generation-level) cuts.}
    \label{MTMET}
\end{figure}

 As a pre-selection, we require a single lepton and at least two $b$-jets, as shown in Tab. \ref{pre-selection}. We use a working point, where the $b$-jet tagger achieves $70\%$ efficiency and only a $1.5\%$  probability of misidentifying a light-parton jet as a $b$-one \cite{CMS:2012feb}.  Then several cuts are imposed to select the Higgs signal as per the  process in Fig. \ref{fig:seesaw3}. The leading lepton is dominantly produced from the process $\tilde\nu_1\rightarrow e + \tilde{\chi}_{1}^{\pm}$. As the mass difference between sneutrino and the lightest chargino is larger than 500 GeV for BP1 and 400 GeV for BP2, we choose the transverse momentum of the leading lepton to be larger than 400 GeV to preserve the single lepton signal and reduce the background, as shown in Fig. \ref{PTL1}. The $\slashed{E}_T$ (MET) cut is chosen to be 500 GeV as the NLSP mass is around that value. In order handle properly the MC generation of the $t\bar t$ background, we add a cut at the generation level (MET above $300$~GeV) so as to generate this SM process automatically in the signal region of interest.  The Higgs selection is done by choosing the interval of invariant mass of the leading and next-to-leading $b$-jets from 100 GeV to 150 GeV. Fig. \ref{IMbb} shows a peak around the SM-like Higgs mass for the signal and $W^\pm h$ background, while the $t\bar t$ noise is rather flat therein. Hence, this requirement proves effective against the latter. Finally, the 100 GeV cut on the transverse mass defined using the highest $p_T$ lepton
 plus missing transverse momentum, $M_T(l_1,{\slashed{E}}_T)$,  can also significantly reduce  background, especially $t\bar t$, as evident from   Fig.~\ref{MTMET}. 

\begin{table}[h]
    \centering
    \begin{tabular}{|c|c|}
    \hline
    Transverse momentum & \\
 of leading lepton &  $ p_T(\ell_1) > 400$~GeV \\
    \hline
    Missing transverse energy   &  $\slashed{E}_T >$ 500 GeV\\
    \hline
    Invariant mass of $b_1$, $b_2$ & 100 GeV $<M(b_1 b_2)<$ 150 GeV \\
    \hline
    Transverse mass & \\
    of leading lepton & $M_T(\ell_1, \slashed{E}_{T})>$ 100 GeV \\
    and missing momentum &  \\
    \hline
    \end{tabular}
    \caption{The full set of cuts used in the MC analysis.}
    \label{cut}
\end{table}

\subsection{Numerical analysis}
We have applied the cuts of Tab.~\ref{cut}  to all BPs as well as backgrounds and the results are presented in Tab.  \ref{cutflow}, for the discussed HE-LHC energy and luminosity. As expected, Type-III seesaw preserves more signal events (25.8 for BP1 and 27.7 for BP2) than Type-I seesaw (15.5 for BP3 and 9.2 for BP4). Furthermore, BP2 and BP4 show the interesting feature of having fewer initial events (compared to BP1 and BP3, respectively) but displaying a similar final result. This is because the sneutrino and smuon in BP2(BP4) are heavier than those in BP1(BP3), leading to a larger MET and higher transverse momentum of the leading lepton ($p_T(\ell_1)$), thereby increasing the efficiency of the corresponding selections. 

The significances are shown in Tab. \ref{Sig}, for the usual HE-LHC parameters,  wherein one can appreciate rather significant signal excesses above the SM backgrounds for Type-III seesaw while for Type-I seesaw the sensitivity is somewhat limited (but larger values of Yukawa couplings could be probed and there could be room to improve the analysis or increase the amount of data). We also tested a benchmark similar to BP1, but with the mass ordering $m(\tilde{e})<m(\tilde{\tau})<m(\tilde{\mu})$ with the smuon too heavy to be produced. This gave just $0.6$ events after the cuts, so we can get a significant signal only arising from selectrons or smuons and their sneutrinos.

\begin{table*}[h]
    \centering
    \hspace*{-1.75truecm}
    \begin{tabular}{|c|c c c c|c c c|} 
    \hline
      Cut  &  BP1 & BP2 & BP3 & BP4 & $W^\pm h$ & $t\bar t$ & Total background \\
      \hline
      Initial &  6150 & 3650 & 6220 & 3660 & 9250000 & 2170000 & 11420000 \\
      $N(b) \geq 2$ & 1095 & 549 & 506 & 193 & 1585690 & 691090 & 2276780 \\
      $N(\ell)=1$ & 482 & 255 & 238 & 90.4 & 835311 & 224927 & 1060238 \\
      \hline
      $\slashed{E}_T >$ 500 GeV & 136.5 & 105.1 & 60.3 & 37.3 & 423 & 126239 & 126662 \\
      $M(b_1b_2)>$ 100 GeV & 113.2 & 86.5 & 52.9 & 27.1 & 380 & 118569 & 118949 \\
      $M(b_1b_2)<$ 150 GeV & 46.7 & 43.8 & 28.0 & 13.9 & 236 & 10706 & 10942 \\
      $ p_T(\ell_1) > 400$~GeV & 25.8 & 27.7 & 15.5 & 9.2 & 15.3 & 203 & 218 \\
   $M_T(\ell_1, \slashed{E}_{T})>$  100 GeV & 25.8 & 27.7 & 15.5 & 9.2 & 3.1 & 19.9 & 23.0 \\
    \hline
    \end{tabular}
    \caption{The response of the signal BPs and  backgrounds to the application of the full cutflow used in the MC analysis. The luminosity is 10 ab$^{-1}$ and the energy is $\sqrt{s}=27$~TeV.}
    \label{cutflow}
\end{table*}

\begin{table}[h]
    \centering
    \begin{tabular}{|c|c|c|c|c|}
    \hline
         & BP1 & BP2 & BP3 & BP4   \\ \hline
    Significance  & $3.69\sigma$ & $3.89\sigma$ & $2.49\sigma$ & $1.62\sigma$\\
    \hline
    \end{tabular}
    \caption{Significance of the BPs for the HE-LHC parameters of Tab.~\ref{cutflow}.}
    \label{Sig}
\end{table}

In addition it is essential for our analysis that there is a significant mass splitting between the sleptons and the LSP. With a softer MET cut the $t\overline{t}$ background would be problematic, while the cut on the transverse mass of the lepton and MET would keep $W^{\pm}h$ under control.

In summary, though, it is clear that the HE-LHC is 
a machine with clear potential to access high scale seesaw models (like Type-III and Type-I embedded within the MSSM) by exploiting the SM-like Higgs (eventually decaying to $b\bar b$) plus a hard lepton and MET signature.

\section{Conclusions}

How neutrino mass generation occurs in Nature is one of the outstanding questions in particle physics. Current probes of neutrinos hardly include colliders, as herein such particles appear as ${\slashed{E}}_T$, thereby offering no scope to identify their properties. However, in a supersymmetric world, there exist sneutrinos, which share with neutrinos their interactions. Therefore, given that sneutrinos can decay visibly at the LHC ({\it i.e.}, inside the detectors), it makes sense, in order to study neutrino properties in supersymmetry, to study sneutrinos. One, however, needs a paradigm for supersymmetry to do so, {\it i.e.},  a model realisation of it, which we assumed here to be the MSSM, supplemented with two kinds of seesaw mechanism for (s)neutrino mass generation, the so-called Type-I and Type-III. These mechanisms have a similar structure to generate neutrino masses and hence both lead to Higgs-sneutrino interactions, which are non-diagonal in flavour space. 

These two are examples of high scale seesaw mechanisms, wherein the companion neutrinos (to the SM ones) can have masses of order $10^{12}-10^{14}$ GeV. However, left-handed sneutrino and slepton masses are necessarily linked to the typical supersymmetry breaking scale, which ought to be 10 TeV or so at the most (in order to preserve gauge coupling unification, successful dynamical EWSB, etc.). In the case of a high seesaw scale the neutrino Yukawa couplings are among the largest ones in the model and, due to the structure of the supersymmetric scalar potential, they can lead to observable consequences at the supersymmetry breaking scale. We found that the current LHC, for which $\sqrt s=14$ TeV (in turn recalling  that $\sqrt{\hat s}$ is only a fraction of that), cannot test such seesaw scenarios. However, a possible energy upgrade has been proposed for it: the so-called HE-LHC. This offers $\sqrt s=27$ TeV (and $\int L\, dt=10$ ab$^{-1}$), therefore, it is in a position to test the aforementioned seesaw scenarios of neutrino mass generation. 

In this paper, we have, in particular, tested the scope of a particular signal stemming from these two seesaw mechanisms. In fact, the signature is common to both, {\it i.e.},  charged current induced slepton-sneutrino production and subsequent decay into the SM-like Higgs boson (in turn decaying to  $b\bar b$ pairs), a single lepton ($l=e,\mu$) and MET (or ${\slashed{E}}_T$). Upon assessing that the single lepton channel (as opposed to multi-lepton ones also stemming in these two scenarios) is the most sensitive one, for any number of $b$-jets beyond 1, 
we have devised a simple cut-and-count analysis, deployed identically for both Type-I and -III, that has enabled us to reach evidence to discovery significances at the HE-LHC for the Type-III case while for the Type-I case a more refined selection and/or additional data would be required. This was shown, in both cases, for BPs currently compliant with standard theoretical requirements as well as current experimental searches.

Parameterwise, the signature requires the gauginos to be heavier than the sleptons, a sufficient mass splitting ($\gtrsim 300$~GeV) between the sleptons and the higgsino-like LSP and a sufficient mass splitting between the slepton generations so that the decay with a Higgs boson is kinematically allowed.

Even though this signal is common to the two seesaw models, the fact that in Type-I seesaw only sneutrinos have decay modes containing Higgs bosons, while for Type-III also charged sleptons have such decay channels allows us to distinguish the models. This distinction might be more difficult at a hadron collider but, if there was an electron-positron collider with sufficient collision energy, the pair production of charged sleptons above $\sqrt{s}=2m_{\tilde{\ell}}$ would lead to an enhanced signal with Higgs bosons in case of Type-III, while no such an enhancement would be present in Type-I.

As an outlook of our work, we would like to highlight that a Future Circular Collider  in hadron-hadron mode (FCC-hh) \cite{FCC:2018byv}, running at $\sqrt s$ values up to 100 TeV, will not improve the scope of the HE-LHC since, herein, background rates increase more that the signal ones that we pursued (although this may not be true for other channels not considered here).

Altogether, we have shown that there exist cases where, in supersymmetric theories, it is possible to probe the neutrino mass generation mechanism through sneutrino phy\-sics while the (seesaw) scale related to this mechanism is extremely high, roughly, up to $10^{14}$~GeV.

\section*{Acknowledgements}
SM is  supported in part through the
NExT Institute and STFC Consolidated Grant No. ST/L000296/1. 
HW is supported by the Carl Trygger Foundation under grant No. CTS18:164.
We finally acknowledge the use of the IRIDIS5 High-Perfor\-mance Computing Facility and associated
support services at the University of Southampton in the completion of this work.

\end{document}